\documentclass[iop]{emulateapj}

\usepackage{graphicx}
\usepackage{amsmath,amssymb}
\usepackage{afterpage}
\usepackage{soul}
\usepackage{color}
\usepackage{url}

\def\la{\mathrel{\hbox{\rlap{\hbox{\lower4pt\hbox{$\sim$}}}{\raise2pt\
hbox{$<$}}
}}}
\def\ga{\mathrel{\hbox{\rlap{\hbox{\lower4pt\hbox{$\sim$}}}{\raise2pt\
hbox{$>$}}
}}}
\def\ergs{~{\rm erg~s^{-1}}}
\def\Msun{\rm{~M}_{\odot}}

\shorttitle{X-ray spectral residuals in NGC 5408 X-1}
\shortauthors{A. D. Sutton et al.}

\begin{document}

\title{X-ray spectral residuals in NGC 5408 X-1: diffuse emission from 
star formation, or the signature of a super-Eddington wind?}

\author{Andrew D Sutton}
\affil{Astrophysics Office, NASA Marshall Space Flight Center, ZP12, Huntsville, Al 35812, USA}
\affil{Department of Physics, University of Durham, South Road, Durham, DH1 3LE, UK}
\email{andrew.d.sutton@nasa.gov}
\author{Timothy P. Roberts}
\affil{Centre for Extragalactic Astronomy, Department of Physics, University of Durham, South Road, Durham, DH1 3LE, UK}
\author{Matthew J Middleton}
\affil{Institute of Astronomy, University of Cambridge, Madingley Road, Cambridge, CB3 0HA, UK}

\begin{abstract}
If ultraluminous X-ray sources (ULXs) are powered by accretion onto stellar 
remnant black holes, then many must be accreting at super-Eddington rates. It
is predicted that such high accretion rates should give rise to massive, radiatively-driven winds. 
However, observational evidence of a wind, in the form of absorption or emission 
features, has remained elusive. As such, the reported detection of X-ray 
spectral residuals in {\it XMM-Newton} spectra of NGC 5408 X-1, which could be 
related to absorption in a wind is potentially very exciting. However, it has 
previously been assumed by several authors that these features simply originate 
from background diffuse plasma emission related to star-formation in the ULX's host galaxy. In this work 
we utilise the spatial resolving power of {\it Chandra} to test whether we can 
rule out this latter interpretation. 
We demonstrate that the majority of the 
luminosity in these spectral features is emitted from a highly localised region close to the ULX, 
and appears point-like even with {\it Chandra}. It is therefore highly likely 
that the spectral features are associated with the ULX itself, and little of 
the flux in this spectral component originates from spatially extended emission in the host galaxy. This may be consistent with the 
suggestion of absorption in an optically thin phase of a super-Eddington wind. 
Alternatively, we could be seeing emission from collisionally ionised material 
close to the black hole, but critically this would be difficult to reconcile 
with models where the source inclination largely determines the observed X-ray 
spectral and timing properties.
\end{abstract}


\section{Introduction}

Whilst it was previously suggested that ULXs could contain intermediate mass 
black holes \citep{colbert_and_mushotzky_1999}, several well-developed arguments have been 
put forward to dispute this (e.g. \citealt{grimm_etal_2003,king_2004}). 
Instead, many ULXs may actually be powered by accretion onto stellar-remnant 
black holes, which has been confirmed in a few cases (e.g. 
\citealt{liu_etal_2013,middleton_etal_2013,motch_etal_2014}; although at least 
one ULX contains a neutron star, see \citealt{bachetti_etal_2014}). Some of these may be 
more massive than the typical $\sim 10 \Msun$ stellar-mass black holes that we see throughout our Galaxy, 
as stellar collapse in regions of low metallicity could 
leave remnants with masses of up to $\sim 80 \Msun$ 
\citep{zampieri_and_roberts_2009, mapelli_etal_2010}. 
However, the Eddington luminosity of even an $80 \Msun$ black hole 
is only $\sim 10^{40}~\ergs$, and ULXs are observed to exceed this. 
Hence, if the majority of ULXs do contain stellar-remnant black holes, then many must be accreting 
matter at close to or in excess of the Eddington limit.
It is therefore important to note that high quality ULX X-ray spectra are observed to differ 
from the standard sub-Eddington states, with many displaying 
both a soft excess and high energy curvature (e.g. 
\citealt{stobbart_etal_2006, gladstone_etal_2009, bachetti_etal_2013}). 

Whilst the X-ray spectra from ULXs below $\sim 3 \times 10^{39}~\ergs$ are typically broad, disc-like continua, 
those from more luminous sources have a characteristic two component shape. 
These peak in either a hard or soft component, in what are termed 
the hard and soft ultraluminous regimes respectively \citep{sutton_etal_2013b}. 
The hard component may 
originate in a cool, optically thick corona \citep{gladstone_etal_2009}, or the 
hot inner disc itself, with a large colour correction \citep{kajava_etal_2012}. 
The soft emission may originate from the base of a radiatively 
driven wind, which is predicted to arise at super-Eddington accretion rates 
(\citealt{shakura_and_sunyaev_1973, poutanen_etal_2007, kajava_and_poutanen_2009}; but see 
\citealt{miller_etal_2013}). Although it has previously been mooted that the 
distinction between these two regimes is largely driven by accretion rate, 
recent work suggests that the inclination of the ULX is also key 
\citep{sutton_etal_2013b, middleton_etal_2015}. 
One potential physical mechanism for introducing the inclination dependence is the
super-Eddington outflow, which is expected to be funnel-shaped \citep{shakura_and_sunyaev_1973,kawashima_etal_2012}. Whilst the soft emission is 
relatively isotropic, the hard emission from the centre of the accretion disc can be geometrically 
beamed by the wind funnel \citep{king_2009}. Hence, when viewed close to face-on the hard emission is focussed towards the observer's line-of-sight and the ULX appears with a hard ultraluminous spectrum. However, at higher 
inclinations the hard emission is scattered away from the line-of-sight and the source appears with a soft ultraluminous spectrum.
This interpretation is 
supported by X-ray timing evidence \citep{sutton_etal_2013b}, if the edge of the wind 
is clumpy at its base \citep{middleton_etal_2011a}, as is the case in 
radiation-magnetohydrodynamic simulations \citep{takeuchi_etal_2013}.

One potential test of the funnel-shaped wind model is a search for the signatures of the
outflow in ULX X-ray spectra, in the form of absorption and emission lines. 
In this vein, \cite{walton_etal_2012, walton_etal_2013} examined 
the Fe K band in NGC 1313 X-1 and Ho IX X-1, finding that any narrow atomic 
absorption or emission features must be intrinsically weak or absent. 
However, both NGC 1313 X-1 and Ho IX X-1 typically have hard ultraluminous X-ray spectra.\footnote{One of the {\it XMM-Newton} observations of NGC 1313 X-1 analysed by \cite{walton_etal_2013}
was classified as soft ultraluminous and a second as ambiguous (either hard of soft ultraluminous) by \cite{sutton_etal_2013b}. However, out of a total of 164 ks of good time in the EPIC pn detector 
analysed by \cite{walton_etal_2013}, these observations only contributed 9 and 10 ks respectively. A
further observation with 3 ks of good EPIC pn time was not classified by \cite{sutton_etal_2013b}. As such, the 
vast majority of the NGC 1313 X-1 data studied by \cite{walton_etal_2013} comes from epochs when the source had a hard ultraluminous spectrum.}
In the funnel-shaped wind model the absorbing material would predominantly intercept 
an observer's line-of-sight in soft ultraluminous sources, and we would not expect 
as strong absorption features to be detected in hard ultraluminous sources. 
Instead, to test the model we ideally want to search for atomic features in soft 
ultraluminous ULXs, but the requirement for a 
high count rate makes such an experiment infeasible in the Fe K band in these 
sources. However, it is potentially very interesting that some soft 
ultraluminous ULXs do show spectral residuals to their best-fitting continuum 
models (e.g. \citealt{stobbart_etal_2006}), including the object that we 
scrutinise in this paper: NGC 5408 X-1 (2XMM J140319.6$-$412258; e.g. 
\citealt{strohmayer_and_mushotzky_2009, miller_etal_2013, middleton_etal_2014}).

At a distance of 4.8 Mpc \citep{karachentsev_etal_2002}, NGC 5408 X-1 is one of 
the best studied ULXs. It has been observed by several of the current generation 
of X-ray satellites, on a multitude of occasions. Observations include: an {\it 
XMM-Newton} large programme (e.g. \citealt{pasham_and_strohmayer_2012}); {\it 
Swift} XRT monitoring \citep{kaaret_and_feng_2009, grise_etal_2013}; and 8 {\it 
Chandra} exposures, which we re-analyse here. 
The flux variability of the source rules out an X-ray supernova remnant and confirms that 
it is powered by accretion onto a compact object \citep{kaaret_etal_2003, soria_etal_2004}.
It persistently displays a distinct soft 
ultraluminous two component X-ray spectrum in {\it XMM-Newton} data 
\citep{sutton_etal_2013b} at an average 0.3--10 keV unabsorbed luminosity of $\sim 1.1 \times 
10^{40} \ergs$ 
(\citealt{strohmayer_2009}; although we note 
that they fit the high energy spectrum with a soft power-law, so may over-estimate 
the intrinsic luminosity, cf. \citealt{middleton_etal_2014}). 
Additional soft 
residuals have been detected in the {\it XMM-Newton} spectra which can be well modelled as 
thermal plasma emission \citep{strohmayer_and_mushotzky_2009, miller_etal_2013, middleton_etal_2014}. It has previously been assumed that these were the result of diffuse 
star-formation related emission in the host galaxy \citep{strohmayer_etal_2007, 
strohmayer_and_mushotzky_2009, miller_etal_2013}. However, 
we know from observational studies of galaxies that the X-ray luminosity of such emission is 
correlated with star-formation rate ($L_{0.5-2 {\rm keV}} / SFR \approx  
8.3 \times 10^{38} \ergs~(M_{\odot}~{\rm yr}^{-1})^{-1}$; 
\citealt{mineo_etal_2012}), and \cite{middleton_etal_2014} contend that the 
luminosity of the putative thermal plasma emission ($\sim 2.5 \times 10^{38}~{\rm erg~s^{-1}}$, 
calculated from \citealt{miller_etal_2013}) greatly exceeds that inferred 
from star-formation, even over the entirety of NGC 5408 ($\approx 3 \times 
10^{37} \ergs$, calculated from a 24 $\rm \mu m$ flux density of $0.42 \pm 0.04~{\rm Jy}$, \citealt{dale_etal_2005}). Instead, \cite{middleton_etal_2014} show that the 
putative plasma emission features could actually be commensurate with broadened, blueshifted absorption in a 
partially ionised, optically thin medium. Such a medium would be expected to 
occur in a super-Eddington wind, as it becomes optically thin at large distances 
($> 10^3 R_{\rm g}$) from the central black hole.

In this paper, we utilise {\it Chandra} archival data 
to determine the spatial origin of the soft plasma emission-like features in the 
X-ray spectrum of NGC 5408 X-1. 
The high angular resolution of {\it Chandra} is absolutely key to this analysis. 
Having constrained the spatial origin of these spectral features, we are able to make 
important inferences as to their nature. In section 
\ref{results} we examine the spatial and spectral properties of the ULX and 
surrounding regions in {\it Chandra} ACIS data, and interpret these results in 
section \ref{discussion}.

\section{Analysis and results}\label{results}

\begin{figure*}
\begin{center}
\includegraphics[width=12cm]{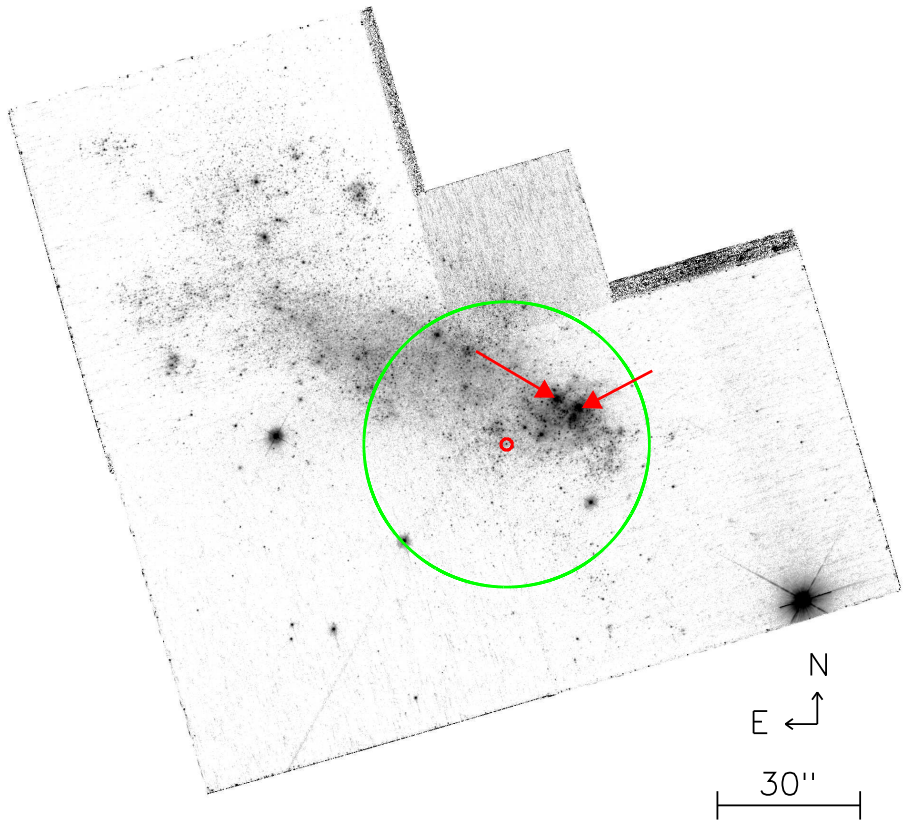}
\caption{{\it HST} WFPC2 image in the F336W near-UV filter of NGC 5408, 
downloaded from the Hubble Legacy Archive. The position of NGC 5408 X-1, taken 
from the {\it Chandra} Source Catalog \citep{evans_etal_2010}, is indicated by 
the arbitrarily-sized red circle, whilst the larger green circle corresponds to 
a nominal 30 arcsec {\it XMM-Newton} source extraction region centred on the ULX. 
Clearly the ULX is not embedded in a young 
stellar association, where star-formation-related diffuse emission would be 
expected. Instead, it is located about $\sim 12$ arcsec from the major regions 
of star-formation in the galaxy (\citealt{kaaret_etal_2003}; red arrows). 
Diffuse emission from these regions may contaminate the {\it XMM-Newton} 
spectrum of NGC 5408 X-1, but can trivially be spatially resolved from the ULX 
by {\it Chandra} ACIS.}
\label{hst}
\end{center}
\end{figure*}

Before examining the {\it Chandra} X-ray observations of NGC 5408 X-1, we 
briefly consider the optical/UV data. NGC 5408 X-1 is located around $\sim 12$ 
arcsec from the major regions of star-formation in NGC 5408 (Figure \ref{hst}; 
\citealt{kaaret_etal_2003}). This would fall well within a typical {\it XMM-Newton} 
source extraction region centred on the ULX, thus star-formation-related emission could feasibly contaminate the 
X-ray spectrum. However, NGC 5408 X-1 is sufficiently displaced from these star-formation regions that 
{\it Chandra} ACIS can spatially resolve them from the ULX. 
There is a much smaller stellar association containing $\sim 20$ OB 
stars, located around 4 arcsec North East of NGC 5408 X-1 
\citep{grise_etal_2012}, and this may still slightly contaminate the {\it Chandra} 
spectrum of the source.

\begin{table}
\centering
\caption{{\it Chandra} ACIS-S observations of NGC 5408 X-1}
\begin{tabular}{ccccc}
\hline
Obs ID & Date         & Exposure time \\
        & (yyyy-mm-dd) & (ks) \\
\hline
 4555 & 2003-12-20     &  5.2 \\
 4556 & 2004-02-09     &  4.9 \\
 4557 & 2004-12-20     &  5.1 \\
 4558 & 2005-01-29     &  5.2 \\
11032 & 2010-05-02     & 12.2 \\
11033 & 2010-05-15     & 12.2 \\
11034 & 2010-05-28     & 12.0 \\
13018 & 2010-09-12     & 12.0 \\
\hline
\end{tabular}
\label{obs}
\begin{minipage}{\linewidth}
Details of the {\it Chandra} ACIS-S observations of NGC 5408 X-1 taken with the 
1/8 sub-array.
\end{minipage}
\end{table}

\begin{figure}
\begin{center}
\includegraphics[width=8cm]{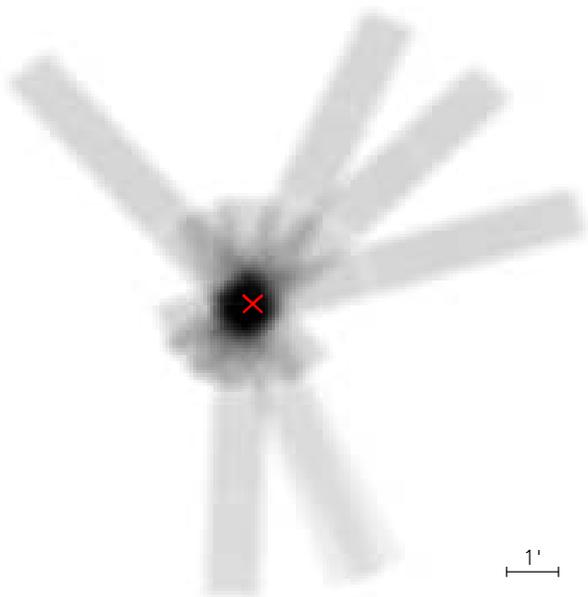}
\caption{Exposure map of the merged {\it Chandra} observations. The approximate 
location of NGC 5408 X-1 is indicated by the red cross. As the ULX is close to 
the nominal ACIS-S aim-point in all of the observations, their individual 
footprints all overlap at the position of the target.}
\label{expmap}
\end{center}
\end{figure}

The region around NGC 5408 has been observed with {\it Chandra} on a total of 9 
occasions, 8 of which are suitable for studying X-1. In this work, we analyse these 8 observations, 
which were taken 
using the 1/8 sub-array of ACIS-S, with X-1 being positioned close to on-axis. In the case 
of NGC 5408 X-1, the use of the 1/8 sub-array is critical to mitigate the 
effects of pileup. 
We estimate residual pileup fractions of $\sim 5$ per cent based on {\sc PIMMS} (v4.7b: 
with ACIS Pile up and Background Count 
Estimation)\footnote{\url{http://cxc.harvard.edu/toolkit/pimms.jsp}} simulations of an 
absorbed power-law with parameters from \cite{gladstone_etal_2009} and an unabsorbed 
0.3--10 keV luminosity of $1.1 \times 10^{40}~\ergs$ \citep{strohmayer_2009}. 
Details of the observations included in this work are given 
in Table \ref{obs}, and a combined exposure map is shown in Figure \ref{expmap}.
We downloaded each dataset from the HEASARC 
archive,\footnote{\url{http://heasarc.gsfc.nasa.gov}} and analysed them using 
tools in {\sc ciao} 4.6,\footnote{\url{http://cxc.harvard.edu/ciao/}} with 
calibration database 4.6.3. We note that we would not expect the analysis to 
change significantly if {\sc ciao} 4.7 was used instead.

\begin{figure}
\begin{center}
\includegraphics[width=8cm]{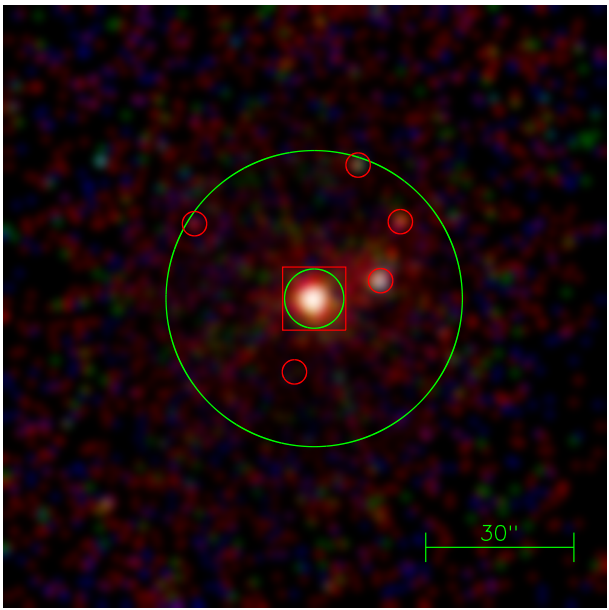}
\caption{Three colour {\it Chandra} ACIS-S image of NGC 5408 X-1 and the surrounding regions, centred on the ULX. The 
RGB colours correspond to 0.2-1.5 keV, 1.5-2.5 keV and 
2.5-8.0 keV respectively. To create the image, all eight observations reported 
in Table \ref{obs} were stacked, and the resulting image was divided by the 
total exposure map (with 8$\times$8 pixel binning), before being convolved with a 2.4 
arcsec FWHM Gaussian to smooth the final image. The green annulus shows the 
6--30 arcsec region in which we estimated the extended emission, and the red box corresponds 
to the footprint of Figure \ref{arestore}. The smaller, arbitrarily sized red circles show the location of other 
sources detected by {\sc wavdetect} within 30 arcsec of X-1.}
\label{chan_img}
\end{center}
\end{figure}

The first stage of the analysis was to confirm that the ULX appeared point-like 
at the spatial resolution of {\it Chandra}. To do this, the individual 
observations were reprocessed using the {\sc chandra\_repro} script, before 
being stacked using {\sc merge\_obs}.  
The stacked image is shown in Figure \ref{chan_img}, where it has been divided 
by the total exposure map of the {\it Chandra} observations (with $8 \times 8$ 
pixel binning). A number 
of faint point sources are evident, which would be confused with the ULX in typical 
30 arcsec {\it XMM-Newton} source extraction regions.

We applied 
the {\sc src\_extent} script to the individual observations to determine whether NGC 5408 X-1 
is extended. In preparation for 
this, we simulated the {\it Chandra} ACIS-S point spread function (PSF) for each 
observation using {\sc 
chart} ({\it Chandra} ray tracer)\footnote{\url{http://cxc.harvard.edu/chart/}} and the {\sc marx} (model of {\it AXAF} 
response to X-rays version 5.0.0; \citealt{davis_etal_2012}) 
software suite. As inputs to {\sc chart}, we used the off-axis angles, azimuthal 
angles and exposure times from each observation, along with absorbed power-law 
model spectra. These model spectra were fitted to the point source spectra (see below for 
details of the spectral extraction) in the 0.3--10 keV energy range using {\sc 
sherpa}, and extracted in an appropriate format using the {\sc 
save\_chart\_spectrum} tool from the {\sc chart\_spectrum} contributed {\sc 
sherpa} package. The simulated PSFs and the reprocessed event files from each 
observation were used as inputs to {\sc src\_extent}. This script estimates the 
size of a source using the Mexican Hat Optimization algorithm, and compares this 
with a PSF. In all but one case, this analysis indicated that the source was 
consistent with having zero intrinsic size. The sole exception to this was 
observation 11034, which had an estimated intrinsic size of $0.14^{+0.05}_{-0.04}$ 
arcsec (which is not flagged as extended by {\sc srcextent} at the default 
threshold criteria, and we note that the PSF and observed source sizes are consistent 
at the 90 per cent significance level).

\begin{figure}
\begin{center}
\setlength{\fboxsep}{1pt}
\setlength{\fboxrule}{1pt}
\fbox{\includegraphics[height=8cm]{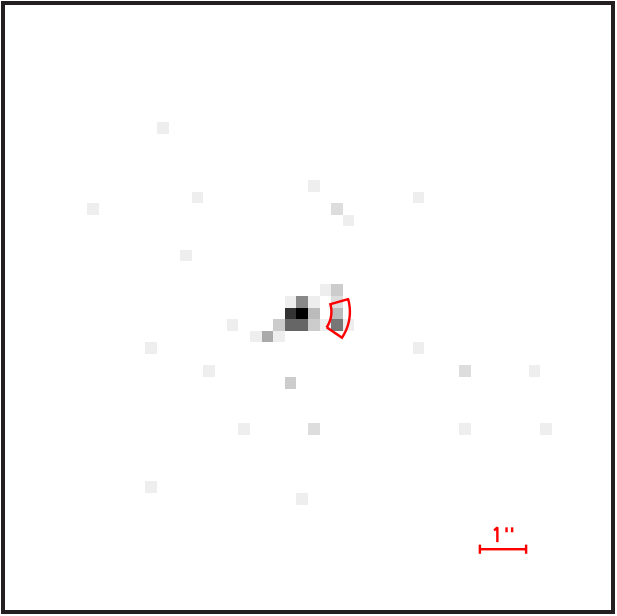}}
\caption{{\it Chandra} ACIS-S 0.5--8 keV image centred on NGC 5408 X-1 from 
observation 11034. The displayed image is $26 \times 26$ native {\it Chandra} 
ACIS pixels in size, and has been rebinned to 0.5 pixels. It has been restored 
using the {\sc arestore} tool in {\sc ciao} with a simulated PSF and 100 
iterations of the Lucy-Richardson deconvolution algorithm. The region 
highlighted in red corresponds to a known asymmetry in the {\it Chandra} PSF, so 
the excess of counts here is unlikely to be a real source. Apart from the ULX 
itself and this PSF asymmetry, no sources were detected by {\sc wavdetect} with 
greater than 10 counts.}
\label{arestore}
\end{center}
\end{figure}

In order to further examine the potential extended emission in observation 11034 
on finer scales, we used {\sc arestore} in {\sc ciao} to deconvolve the observed 
data with a simulated PSF. For this purpose, we created a PSF using {\sc chart} 
and {\sc marx} with 10 times the exposure time of the real observation. The {\sc 
chart} raytraces are deterministic rather than statistical, so increasing the number of photons 
improves our sampling of the PSF. Images of both the observed data and the 
simulated PSF were created with 0.5 pixel binning. These were used as inputs to 
{\sc arestore}, which restores images which have been degraded by a blurring 
function using the Lucy-Richardson deconvolution algorithm 
\citep{richardson_1972, lucy_1974}. We ran the deconvolution algorithm for 100 
iterations, after which some additional structure was evident around $\sim$ 0.7 
arcsec to the West of the ULX (Figure \ref{arestore}). 
However, this region corresponds to a known asymmetry fixed in spacecraft 
coordinates,\footnote{\url{http://cxc.harvard.edu/ciao/caveats/psf\_artifact.html
}} and the excess of counts is highly likely to be this artefact. 
To test for any other significant resolved sources we ran {\sc wavdetect} on the restored 
image, with wavelet radii of 1, 2 and 4 image pixels. Apart from the ULX itself and 
the known artefact, no other sources with greater than 10 counts were detected in the region shown in Figure \ref{arestore}. 
As such, we find no convincing evidence that the ULX can be further spatially 
resolved by {\it Chandra} ACIS.

In addition to the spatial analysis, we also looked for an excess of spectral 
counts in the regions surrounding the {\it Chandra} point source, to further test whether the 
putative plasma emission component could originate there. 
We fit this emission with a variety of models, a summary of which is given in 
Table \ref{extended_fits}. 
We note that the nature of the features
in the {\it XMM-Newton} spectrum is 
not clear, and they may originate from either emission or absorption. Throughout the rest of this work we model them as 
plasma emission, but do not intend to imply that this is the correct physical interpretation.
Since the response of {\it Chandra} ACIS changed significantly during the 7 year time span over which the 
observations were taken, especially at low energies, it is not acceptable to simply extract a merged spectrum.
As such, we used {\sc 
specextract} in {\sc ciao} to extract energy spectra from each observation in a 
6--30 arcsec annular region centred on NGC 5408 X-1 (Figure \ref{chan_img}), with appropriate parameters 
for an extended source ({\sc correctpsf}=no and {\sc weight}=yes). 
It is clear from Figure 3 that several point sources were 
resolved by {\it Chandra} within $\sim 30$ arcsec of NGC 5408 X-1. As such, we used 
{\sc wavdetect} with {\sc scales}=`1, 2, 4, 8, 16' and {\sc sigthresh}=$1 \times 10^{-5}$ (i.e. $\sim$ 1 false detection over the number 
of pixels in the image) to identify sources near to X-1.
Sources within 30 arcsec of the ULX are highlighted in Figure 3, and were masked out when defining the extended source region. 
The {\sc 
specextract} script also produces appropriate response matrix files (RMFs), 
ancillary response files (ARFs) and background spectra, which were estimated from 
a large circular region, far from the ULX. 

\begin{table*}
\centering
\caption{Spectral models used to fit the extended emission}
\label{extended_fits}
\begin{tabular}{cccc}
Model$^a$ & E range$^b$ & Mask$^c$ & Goodness-of-fit$^d$ \\
\hline
$\textsc{tbabs} \times \textsc{mekal}$ & 0.3--2 keV & y & 94\% \\
$\textsc{tbabs} \times \textsc{mekal}^e$ & 0.3--2 keV & y & 76\% \\
$\textsc{tbabs} \times \textsc{mekal}$ & 0.3--2 keV & n & 99\% \\
$\textsc{tbabs} \times (\textsc{mekal} + \textsc{tbabs} \times \textsc{powerlaw})$ & 0.3--10 keV & n & 97\% \\
\hline
\end{tabular}
\begin{minipage}{\linewidth}
Summary of the goodness-of-fit statistics for spectral models simultaneously fitted to 
the extended emission from all of the {\it Chandra} observations. All model parameters were 
assumed to be invariable between observations. We note that from a 
comparison of the first and second models 
it seems that fixing a parameter of the model to a non-optimal value actually improves 
the goodness of fit. This is because different fit and test statistics were used.
Notes:
$^a${\sc xspec} model fitted to the extended spectrum;
$^b$energy range in which the model was fitted;
$^c$flag for whether a point source mask was applied when extracting the extended spectrum, to remove contaminating emission from resolved point sources;
$^d$goodness of fit of the model in terms of the percentage of 10000 simulated 
realisations based on the model which had improved Anderson-Darling statistics compared to 
the real data;
$^e$unlike the other models, here the {\sc mekal} temperature is fixed to the value given in {\cite{miller_etal_2013}}.
\end{minipage}
\end{table*}

The unbinned extended source spectra, RMFs, ARFs and background spectra were 
read in to {\sc xspec} (version 12.8.1), and simultaneously fitted with a plasma 
emission model in the 0.3--2 keV energy range. When fitting the model there were insufficient 
data in each observation to allow us to use binning and the $\chi^2$ statistic, so we used 
the modified Cash statistic in {\sc xspec}, which allows for a background 
spectrum to be 
subtracted.\footnote{ \url{
http://heasarc.gsfc.nasa.gov/xanadu/xspec/manual/X\allowbreak S\allowbreak a\allowbreak p\allowbreak p\allowbreak e\allowbreak n\allowbreak d\allowbreak i\allowbreak x\allowbreak Statistics.html} } 
Specifically, the spectral model that we used was $\textsc{tbabs} \times \textsc{mekal}$. The 
absorption component was fixed to the Galactic value ($5.93 \times 10^{20}~{\rm 
cm}^{-2}$; \citealt{dickey_and_lockman_1990})
and we used the solar abundance table of \cite{wilms_etal_2000}. 
As any 
extended emission would not be expected to vary significantly between observations, the {\sc mekal} 
temperature and normalisation were fixed between the spectra. 
To assess the goodness-of-fit, we used the {\sc goodness} command in {\sc xspec} 
to simulate 10000 realisations from the best fitting model, and calculated the 
Anderson-Darling statistic to compare these with the real data. Of the 10000 
realisations, 94 per cent had an improved fit-statistic compared to the real data, indicating that the 
model cannot be rejected at 95 per cent significance (equivalent to $2 \sigma$ 
for Gaussian distributed data). 
This model had a plasma temperature of $0.64 \pm 0.03~{\rm keV}$, which is in 
disagreement with the value of $0.87 \pm 0.02~{\rm keV}$ reported by 
\cite{miller_etal_2013} for a plasma emission spectrum fitted to the residual features in 
{\it XMM-Newton} EPIC data. However, when we repeated the above goodness-of-fit 
test with the {\sc mekal} temperature fixed at 0.87 keV, we found that this variant of 
the model cannot be rejected at high significance either (76 per cent of 10000 
simulated data sets had a lower Anderson-Darling test statistic; note that 
the test statistic differs from the fit statistic). 
We used the {\sc cflux} convolution model in {\sc xspec} to estimate the observed 
0.3--2 keV flux of the extended emission, integrated over the annular region as 
$(2.5 \pm 0.2) \times 10^{-14}~{\rm erg~cm^{-2}~s^{-1}}$. At the distance of NGC 
5408, this is equivalent to a 0.3--2 keV luminosity of $(7.0 \pm 0.5) \times 
10^{37}~{\rm erg~s^{-1}}$. If we renormalise this to account for the difference 
in region size between our annular region with point source masks and a 
30 arcsec circle, then this is increased by $\sim 6$ per cent to $(7.4 \pm 0.5) 
\times 10^{37}~{\rm erg~s^{-1}}$. This is $\sim 1/3$ of the {\sc mekal} flux seen 
in the {\it XMM-Newton} data.

The {\it XMM-Newton} source extraction region does not exclude the faint point sources resolved by {\it Chandra}, 
so we test whether these could be contributing any of the missing {\sc mekal} flux. To do this,
we repeated the above analysis in the same annular region without masking out the point sources. 
An absorbed {\sc mekal} model could not be rejected at the equivalent of $3 \sigma$ significance in the 0.3--2 keV range 
(99 per cent of 10000 simulated spectra had an improved fit statistic). 
We estimated the observed 0.3--2 keV flux using the {\sc cflux} model in {\sc xspec} to be 
$(3.2 \pm 0.2) \times 10^{-14}~{\rm erg~cm^{-2}~s^{-1}}$ integrated over the annulus. After correcting 
for the difference in region size, this is around 20 per cent higher than the previous estimate
with point sources excluded, but it still can only account for $\sim$ 37 per cent of the 
{\it XMM-Newton} {\sc mekal} luminosity. However, when we tried adding a power-law to the spectral model 
(${\textsc{tbabs} \times (\textsc{mekal} + \textsc{tbabs} \times \textsc{powerlaw}})$ in {\sc xspec}, with the model fitted in the 0.3--10 
keV range, and having a rejection probability of 97 per cent) this resulted in a larger 
increase in the estimated 0.3--2 keV observed flux. Most of the difference in observed flux between the 
two models is in the 1--2 keV energy band, and it is likely due to fitting the models over different energy ranges.
For the latter model the total observed flux was 
$(4.7 \pm 0.3) \times 10^{-14}~{\rm erg~cm^{-2}~s^{-1}}$, but only 
$(1.1 \pm 0.4) \times 10^{-14}~{\rm erg~cm^{-2}~s^{-1}}$ originated in the {\sc mekal} component. As such, much of the faint flux 
would have been confused with the continuum spectrum in {\it XMM-Newton}, not the putative plasma emission features. 
Since the {\sc mekal} flux here is lower than our estimate with the point sources excluded, this may 
suggest that there is still some unresolved faint point source emission. 
We therefore take $(7.4 \pm 0.5) \times 10^{37}~{\rm erg~s^{-1}}$ as a conservative estimate of the extended 
contribution to the {\it XMM-Newton} {\sc mekal} component luminosity, but caution that it should 
be considered as an upper-limit.

We also examined point source spectra from the ULX to test whether there was 
evidence of {\sc mekal}-like features, which could indicate that the spectral 
features have a localised origin close to the ULX. A summary of models fitted to the 
ULX spectra is given in Table \ref{ulx_fits}. Again, we used {\sc 
specextract} in {\sc ciao} to extract the source and background spectra, RMFs and ARFs from each of the 
{\it Chandra} observations. Source spectra were extracted from 5 arcsec circular 
regions centred on the ULX, and background spectra from large, circular regions 
located on the same chip, away from the ULX and other point sources. As is 
appropriate for a point source, we used parameters {\sc correctpsf} = yes and 
{\sc weight} = no. The spectra 
and associated files were grouped and binned to a minimum of 20 counts per energy bin by 
the {\sc specextract} script, allowing us to use the $\chi^2$ statistic. 

\begin{table}
\centering
\caption{Spectral models used to fit the ULX}
\label{ulx_fits}
\begin{tabular}{cc}
Model$^a$ & $\chi^2.{\rm dof}$ \\
\hline
$\textsc{tbabs} \times \textsc{tbabs} \times \textsc{powerlaw}$ & 716.7/623 \\
$\textsc{tbabs} \times (\textsc{tbabs} \times \textsc{powerlaw} + \textsc{mekal})$ & 622.4/621 \\
$\textsc{tbabs} \times \textsc{tbabs} \times \textsc{simpl} \times \textsc{diskbb}$ & 628.2/607 \\
$\textsc{tbabs} \times (\textsc{tbabs} \times \textsc{simpl} \times \textsc{diskbb} + \textsc{mekal})$ & 615.5/605 \\
$\textsc{tbabs} \times (\textsc{tbabs} \times \textsc{simpl} \times \textsc{diskbb} + \textsc{mekal})$ & 617.7/606 \\
$\textsc{tbabs} \times \textsc{tbabs} \times (\textsc{simpl} \times \textsc{diskbb} + \textsc{mekal})$ & 613.5/612 \\
\hline
\end{tabular}
\begin{minipage}{\linewidth}
Summary of the goodness-of-fit statistics for a variety of spectral models simultaneously fitted to 
all of the {\it Chandra} ULX spectra. In all of the  models the first absorption component was set equal 
to the Galactic neutral hydrogen column density in the direction of the source. The second absorption component 
was left free to model intrinsic absorption in the source/absorption in the host galaxy. This was free to 
vary between observations, except in the final model where it was assumed to be invariable.
Similarly, ULX components in all of the models ({\sc powerlaw}, {\sc simpl} and {\sc diskbb}) were allowed to 
vary between observations, whilst the {\sc mekal} parameters was assumed to be invariable. 
Notes:
$^a${\sc xspec} model fitted to the extended spectrum.
\end{minipage}
\end{table}

We read all of the grouped {\it Chandra} point source spectra into {\sc xspec} 
and simultaneously fitted them with a simple phenomenological model of emission 
from a ULX, a doubly absorbed power-law ($\textsc{tbabs} \times \textsc{tbabs} \times 
\textsc{powerlaw}$ in {\sc xspec}). The first absorption component was fixed to the 
Galactic value, and the second was left free to model intrinsic absorption in 
the host galaxy and/or the ULX itself. The power-law spectral index and 
normalisation were allowed to vary between observations, to model any spectral 
variability in the ULX. Although it was not strictly necessary to simultaneously 
fit the spectra, we chose to do this to later allow us to assess the 
significance of an additional invariable {\sc mekal} component. This model has a 
goodness-of-fit statistic of $\chi^2/{\rm dof} = 716.7/623$, which cannot 
formally be rejected at $3 \sigma$ significance. Next, we repeat the fitting 
process with an additional {\sc mekal} component, which is kept constant between 
observations ($\textsc{tbabs} \times (\textsc{tbabs} \times \textsc{powerlaw} + \textsc{mekal})$ in {\sc 
xspec}). 
Keeping the {\sc mekal} component fixed is reasonable if it originates in diffuse emission, 
although that is not necessarily the case here where we are looking at the point source.
However, the data are such that we have insufficient spectral counts to constrain a 
variable {\sc mekal} component, so we must make the assumption that it does not vary strongly.
This additional component results in a large improvement in the 
goodness-of-fit over an absorbed power-law alone ($\Delta \chi^2 = 94.3$ for 2 
degrees of freedom). However, the 0.3--2 keV observed luminosity in the {\sc mekal} 
component ($(4.1 \pm 0.4) \times 10^{38}~{\rm erg~s^{-1}}$) is around double 
that seen in the {\it XMM-Newton} EPIC data. Clearly, this is not physically 
realistic, as the putative {\sc mekal} contribution to the {\it XMM-Newton} spectrum 
is here integrated over a much smaller area. Instead, we suggest that as the 
{\it XMM-Newton} spectrum of NGC 5408 X-1 has a two component form (e.g. 
\citealt{gladstone_etal_2009, middleton_etal_2015}), perhaps we are fitting the plasma emission to 
the soft emission component from the ULX itself, not the more 
subtle spectral features. Instead, a two component model of emission from a ULX 
may be more appropriate.

\begin{figure*}
\begin{center}
\includegraphics[width=18cm]{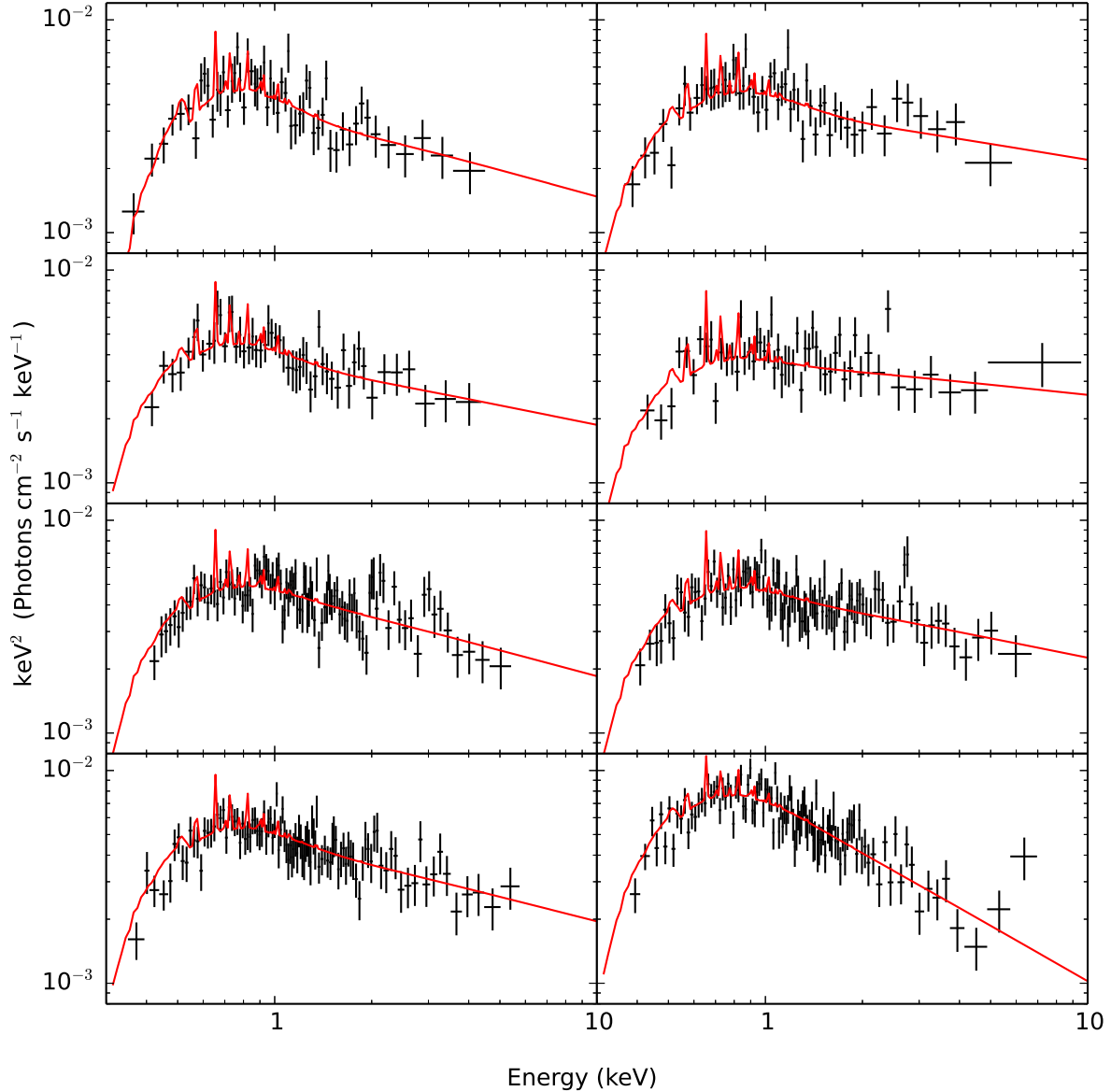}
\caption{X-ray spectra from the eight {\it Chandra} ACIS-S observations detailed 
in Table \ref{obs}. The data have been unfolded using a power-law in {\sc xspec}
with $\Gamma = 0$ and $\rm{Norm} = 1$.
The spectra are shown in order of observation date, starting 
at the top left and reading across the rows. The X-ray spectra were extracted 
using {\sc ciao}, and data points grouped to a minimum of 20 counts per bin, as 
described in the main body of text. The spectral data are over-plotted with the 
best fitting model of a two-component ULX continuum, with an additional 
contribution from diffuse plasma emission ($\textsc{tbab} \times (\textsc{tbabs} \times \textsc{simpl} 
\times \textsc{diskbb} + \textsc{mekal})$ in {\sc xspec}), with the plasma emission component 
fixed between the different observations.}
\label{spec}
\end{center}
\end{figure*}

{\it XMM-Newton} ULX spectra have previously been fitted with models of emission 
from a multi-colour-disc plus Comptonisation in an optically thick medium 
($\textsc{diskbb} + \textsc{comptt}$). 
However, the energy response of {\it Chandra} ACIS is such 
that we could not constrain the high energy curvature, even if it was present. 
So, we instead opted to use a multi-colour-disc with an empirical power-law 
approximation of Comptonisation ($\textsc{tbabs} \times \textsc{tbabs} \times \textsc{simpl} \times 
\textsc{diskbb}$ in {\sc xspec}). One key difference between {\sc simpl} and a power-law is 
that it does not diverge at low energies. This was potentially advantageous 
here, as we were trying to test for the presence of a soft, faint diffuse 
emission component. As previously, we simultaneously fitted the model to all eight {\it 
Chandra} spectra, with parameters free to vary between observations. Again, we did this, 
to allow for a comparison with an additional invariable {\sc mekal} component. This 
results in a goodness-of-fit statistic of $\chi^2/{\rm dof} = 628.2/607$, which cannot be rejected at 2 $\sigma$ significance. From the model, we estimated 
inner disc temperatures of 0.17--0.22 keV and power-law spectral indices of 
typically 2.2--2.4, with the exception of observation 13018, which required a 
softer spectral index of $2.88^{+0.10}_{-0.08}$. 

The addition of the {\sc mekal} component to the two component ULX model (i.e. 
$\textsc{tbabs} \times (\textsc{tbabs} \times \textsc{simpl} \times \textsc{diskbb} + \textsc{mekal})$, Figure \ref{spec})
resulted in a reduction to the fit statistic ($\chi^2/{\rm dof} = 615.5/605$; 
i.e. $\Delta \chi^2 = 12.7$ for 2 degrees of freedom), with a plasma 
temperature of $0.27^{+0.03}_{-0.02}$ keV. 
We note that this additional spectral component is not strongly required by the {\it Chandra} 
data, however we are not testing for its significance here, we are simply constraining how much flux 
it could contain.
Our estimate of the {\sc mekal} temperature is in disagreement 
with the value of $0.87 \pm 0.02$ keV reported by 
\cite{miller_etal_2013}. However, we repeated the spectral fitting with the {\sc mekal} 
temperature fixed at 0.87 keV, and compared the fits using an F-test. Setting the 
{\sc mekal} plasma temperature in this way resulted in an increase in $\chi^2$ of 2.2
for 1 degree of freedom. This corresponds to an F-test probability of 0.14, thus 
the improvement in the fit from allowing the plasma temperature to vary is only 
very marginal ($<2 \sigma$ significance). We estimated the observed 0.3--2 keV 
luminosity of the {\sc mekal} component using the {\sc cflux} convolution model in {\sc 
xspec} to be $L_{\rm X} = 1.3^{+0.7}_{-0.4} \times 10^{38}~\ergs$ (or 
$1.3^{+0.4}_{-0.3} \times 10^{38}~\ergs$ when the {\sc mekal} temperature was fixed to 
0.87 keV).
This luminosity is consistent with the point source providing the observed {\it XMM-Newton} putative {\sc mekal} component that remains after subtracting the spatially resolved diffuse emission.
In this model, the physical interpretation of the second {\sc tbabs} component is not clear, as it 
accounts for both absorption in the host galaxy and intrinsic to the ULX.
Therefore, at least the galactic component of this absorption should be applied to the 
putative plasma emission features, and arguably the intrinsic component too, depending on their physical origin. 
As such, we also tried $\textsc{tbabs} \times \textsc{tbabs} \times (\textsc{simpl} \times \textsc{diskbb} + \textsc{mekal})$, 
with the first absorption component set to the Galactic value and the second free to vary, but fixed between observations. 
In this iteration the mekal temperature was $kT = 0.25 \pm 0.02~{\rm keV}$ and the 0.3--2 keV 
observed component luminosity was slightly higher at $L_{\rm X} = 2.2^{+0.4}_{-0.6} \times 10^{38}~{\rm erg~cm^{-2}~s^{-1}}$, but is still consistent with the previous estimate. 
This plasma temperature is still in disagreement with the value reported by \cite{miller_etal_2013}, but the luminosity is consistent 
with that of the entire {\sc mekal} component seen in {\it XMM-Newton} data. 

\section{Discussion and conclusions}\label{discussion}

Previous {\it XMM-Newton} analyses of NGC 5408 X-1 have identified spectral 
residuals that can be modelled as emission from a thermal plasma. As such, 
it has previously been assumed that these residuals originated in star-formation related 
diffuse emission in the host galaxy, and they were not associated with the ULX itself 
(e.g. \citealt{strohmayer_and_mushotzky_2009, miller_etal_2013}). However, 
\cite{middleton_etal_2014} noted that this would require the diffuse emission within the
{\it XMM-Newton} source extraction region to be in excess of that expected for the entire 
galaxy based on its star-formation rate. Instead, \cite{middleton_etal_2014} 
suggested that the residuals may 
originate in a radiation-pressure-dominated wind coming from the ULX. 
In this work, we have exploited the spatial resolution of {\it Chandra} to place the 
best possible limits on the origin of the putative diffuse plasma 
emission.

From our analysis, we find no strong evidence that the ULX deviates from 
remaining essentially point-like in {\it Chandra} ACIS data. Whilst a hand-full 
of faint point sources not evident in the {\it XMM-Newton} observations are 
resolved, the ULX does not appear to sit within a strongly peaked distribution 
of diffuse emission. 
This is not particularly unexpected, given that the ULX is largely displaced from the 
major star-formation regions in NGC 5408.
This result is also largely confirmed by a spectral 
analysis of the regions surrounding X-1. Although we do find an excess of 
counts within 30 arcsec of the ULX, these are only sufficient to account for a 
0.3--2 keV luminosity of $\lesssim 7.4 \times 10^{37}~{\rm erg~s^{-1}}$. Admittedly, this limit is a 
factor $\sim 2$ greater than the predicted star-formation related luminosity 
integrated over the entirety of NGC 5408 \citep{middleton_etal_2014}, but it may contain some degree of
unresolved point source emission, and is 
still only sufficient to contribute $\sim 1/3$ of the 0.3--2 keV {\sc mekal} flux seen 
in the {\it XMM-Newton} data ($\sim 2.5 \times10^{38}~{\rm erg~s^{-1}}$; 
calculated from \citealt{miller_etal_2013}). 

To test whether the {\it Chandra} 
ULX point source spectra could be harbouring the remainder of the putative plasma 
flux observed in the {\it XMM-Newton} data, we fitted 
them with a model for the continuum spectrum from a ULX, plus an additional {\sc mekal} component. 
The {\sc mekal} component had a 0.3--2 keV luminosity of  $1.3^{+0.7}_{-0.4} \times 
10^{38}~\ergs$ (or $1.3^{+0.4}_{-0.3} \times 10^{38}~\ergs$ assuming a plasma 
temperature of 0.87 keV), thus is consistent with contributing the missing flux. 
We know from the {\it HST} data that a small optical association falls within the 
5 arcsec source extraction region \citep{grise_etal_2012}, 
but {\it Chandra} sees a point source and it seems very unlikely that a 
minor stellar association could produce $\sim 4$ times the 
diffuse X-ray emission than is expected over the entire galaxy. Indeed, we very crudely estimate the 
X-ray luminosity of this association as $\sim 2 \times 10^{33}$--$2 \times 10^{35}~{\rm erg~s^{-1}}$ 
by assuming that all $\sim 20$ of the stars are colliding wind 
binaries \citep{mauerhan_etal_2010}.
We note that the additional {\sc mekal} component is not strongly required by the 
data, although this is hardly surprising given that it contributes only $\sim 
4.6$ per cent of the flux \citep{strohmayer_etal_2007}, and we are much more 
photon-limited here than in the {\it XMM-Newton} observations. Not only does 
{\it XMM-Newton} EPIC have greater effective area, but in the case of NGC 5408 
X-1 the observations also have much greater exposure times: 6 {\it XMM-Newton} 
observations have between 28.6 and 88.2 ks of good time 
\citep{middleton_etal_2014}, compared to at best 12.2 ks with {\it Chandra}. 
However, it is reasonable to include the {\sc mekal} component here, as we have prior 
knowledge that two component phenomenological ULX models excluding it are ruled 
out by the {\it XMM-Newton} data (e.g. \citealt{strohmayer_etal_2007}). 

The majority of the putative plasma flux remains spatially unresolved, and is 
not displaced from the {\it Chandra} point source. Therefore, the putative 
{\sc mekal} features in {\it XMM-Newton} data cannot be dominated by 
star-formation related diffuse emission. Rather, we favour scenarios where they 
are associated with the ULX. Similar 
spectral features have been reported in a number of other ULXs, including: Ho 
II X-1 \citep{miyaji_etal_2001,dewangan_etal_2004}, NGC 4395 X-1 \citep{stobbart_etal_2006}, NGC 4559 X-1 
\citep{roberts_etal_2004}, NGC 6946 X-1 \citep{middleton_etal_2014} and NGC 7424 
ULX2 \citep{soria_etal_2006b}. 
Notably these sources all have soft X-ray spectra, at least in the 
observations where residuals are reported \citep{stobbart_etal_2006, 
sutton_etal_2013b, middleton_etal_2014}.
Furthermore, 
\cite{soria_etal_2006b} reported detections of NGC 7424 ULX2 with both hard and 
soft spectra ($\Gamma = 1.8^{+0.2}_{-0.1}$ and $2.2 \pm 0.2$ respectively 
when fitted with absorbed power-laws), with the plasma-like component only being 
detected when the source had a soft spectrum, thus ruling out the possibility 
that it originates from underlying emission. The fact that we typically see 
these plasma emission-like features in ULXs with soft X-ray spectra strongly suggests 
that they could be associated with a super-critical wind. However, similar features 
have been identified in ULXs with hard X-ray spectra, e.g. NGC 1313 X-1 
\citep{bachetti_etal_2013} and Ho IX X-1 \citep{walton_etal_2014}. This does not necessarily rule out 
an association with an outflow. 
The wind would be expected to become physically diffuse as it moves away from the accretion disc, 
thus the optically thin phase could intercept an observers line-of-sight at low inclinations, even if the optically 
thick material does not.
Additionally,  the wind opening angle is expected to vary with 
accretion rate \citep{king_2008}, so ULXs could shift between hard and soft ultraluminous 
states. Indeed, such shifts have been reported in NGC 1313 X-1 and NGC 5204 X-1 \citep{sutton_etal_2013b}.
Furthermore, Middleton et al. (submitted) show that the strength of the the putative absorption 
features in NGC 1313 X-1 are anti-correlated with spectral hardness, which supports the interpretation 
that they are associated with an outflow.

One possibility, as suggested by \cite{middleton_etal_2014}, is that putative mekal features 
do not originate from plasma emission, rather they are misdiagnosed absorption features from a blue-shifted, 
partially ionised, optically thin absorber. Such a material could originate from 
a radiation-pressure dominated wind driven by super-critical accretion, at large 
distances from the black hole. 
Alternatively, the features may truly be associated with thermal plasma 
emission. This could potentially occur if ejecta in the ULX wind collisionally ionise a 
nearby cloud of material, such as the outer-layers of a highly evolved massive 
stellar companion \citep{roberts_etal_2004}. However, observational data  
\citep{middleton_etal_2011a, sutton_etal_2013b, middleton_etal_2015} and 
simulations \citep{takeuchi_etal_2013} suggest that the distinction between hard 
and soft ultraluminous sources is determined at least in part by the observation angle.
As there is no {\it a priori} reason that such plasma emission would be anisotropic, 
the stronger detections in soft ultralumious ULXs 
is troubling for the collisionally 
ionised plasma emission scenario. 
For this reason, we tend towards favouring absorption as the more likely explanation. 
Future missions with high spectral resolution, such as {\it Astro-H} and {\it Athena}, 
will be able to provide definitive tests of the nature of the features we discuss here. 
However, in the meantime further diagnostics can come from examining their evolution 
with changing X-ray continuum spectra \citep{middleton_etal_2015b}.

\acknowledgments{We thank Douglas Swartz for useful suggestions and discussion.
ADS and TPR acknowledge funding from the Science and Technology
Facilities Council as part of the consolidated grants ST/K000861/1 and
ST/L00075X/1. ADS also acknowledges funding through a NASA Postdoctoral 
Program appointment at Marshall Space Flight Center, administered by 
Oak Ridge Associated Universities on behalf of NASA. MJM appreciates support from ERC grant 340442.}

{\it Facilities}: \facility{CXO}

\bibliography{refs}
\bibliographystyle{apj}

\end{document}